\documentclass{aa}
\usepackage{graphics}
\usepackage{txfonts}
\begin{document}
   \title{A study of bright Southern Long Period Variables}

   \author{T. Lebzelter\inst{1}
          \and
			K.H. Hinkle\inst{2}
			\and
			P.R. Wood\inst{3}
			\and
			R.R. Joyce\inst{2}
			\and
			F.C. Fekel\inst{4}
          }

   \offprints{T. Lebzelter}

   \institute{Institute for Astronomy (IfA), University of Vienna,
              T\"urkenschanzstrasse 17, A-1180 Vienna, Austria\\
              \email{lebzelter@astro.univie.ac.at}
              \and National Optical Astronomy Observatory, P.O. Box 26732, 
Tucson, AZ 85726, USA
              \and Research School for Astronomy and Astrophysics, 
Australian National University, 
              Cotter Road, Weston ACT 2611, Australia
              \and Center for Excellence in Information Systems, 
Tennessee State University,
              330 10th Avenue North, Nashville, TN 37203, USA
             }

   \date{Received ; accepted }

   \abstract{In this paper we present radial velocity curves of AGB variables that exhibit 
   various kinds of anomalies: Semiregular variables 
   (SRVs) with typical mira periods, SRVs exceeding the mira 2.5\,mag 
   amplitude limit, miras with secondary maxima in their light curves, and a SRV with 
   a long secondary period.  The stars with reliable Hipparcos parallaxes 
   from this and from previous studies are plotted in a logP-$M_{K}$-diagram.
   Our objects nicely follow the logP-$M_{K}$-relations determined
   for the LMC. This allows the pulsation mode to be identified.
   While all miras fall on the fundamental mode sequence, the
   SRVs fall on both the first overtone and fundamental mode sequences.
   The SRVs on the fundamental mode sequence occur at both high and low
   luminosities, some of them being more luminous than larger amplitude
   miras.  This demonstrates observationally that some parameter other than luminosity
      affects the stability of long period variables, probably mass.
   First overtone pulsators all show velocity amplitudes around 
   4\,kms$^{-1}$.  For the fundamental mode pulsators, the velocity amplitude shows
   a correlation with light amplitude. The two miras R Cen and R Nor,
   known for their double-peaked light curves, have velocity curves that are
   quite different.  The R Nor velocity curve shows no evidence of the
   double peaks, meaning that the true pulsation period is the time between
   alternate minima or maxima.  There is slight evidence for a double bump in the R Cen
   velocity curve.  It is suggested that these stars are relatively massive (3--5 M$_{\odot}$).\\ 
\\
   }

   \maketitle
%

\section{Introduction}\label{sect1}

Variability due to radial pulsation is a key feature of stars on the Asymptotic
Giant Branch (AGB). These variable stars, collectively know as long-period
variables (LPVs), have generally been put into main groups: miras, semiregular
(SRV) and irregular variables.  The General Catalogue of Variable Stars (GCVS,
Kholopov et al.~\cite{GCVS}) adopts a visual light amplitude of 2.5 mag to separate miras and
semiregular variables.  However, the transition between miras and SRVs is
somewhat uncertain and some stars are found under both classifications in the
literature.

The pulsation velocities of LPVs are best studied in the near-infrared (Hinkle
et al.\,\cite{HHR82}, Wallerstein \cite{Wallerstein85}) as the optical spectrum
gives an inconclusive picture of the stellar pulsation (e.g.\,Wallerstein
\cite{Wallerstein77}).
We have successfully used the near-infrared lines to study pulsation of LPVs
several times in the past (e.g.\,Hinkle \cite{Hinkle78}, Hinkle et
al.\,\cite{HLS97}, \cite{HLJF02}, Lebzelter \cite{L99}, Lebzelter et
al. \cite{LHH99}).  Near infrared velocity data have been obtained for 18 miras
and almost 30 SRVs in the Galactic field and halo. All miras show similar,
sawtooth-shaped velocity curves with amplitudes between 20 and 30\,kms$^{-1}$
and line doubling around maximum light.  SRVs have smaller velocity amplitudes
and continuous, roughly sinusoidal, velocity curves. A recent summary was given
by Lebzelter \& Hinkle (\cite{LH02}). Most recently, we have studied the
velocity variations in 12 LPVs in the globular cluster 47\,Tuc (Lebzelter et
al.\,\cite{LWHJF04}).

The aim of this paper is to contribute to our understanding of the pulsation of
LPVs by studying the near-infrared velocity curves of some stars with unusual
properties.  For our study we selected targets  
showing one or 
several of the following outstanding behaviors:
 
\begin{itemize}

\item {\it SRVs with long periods falling into the domain of the miras 
($\ge$300 days).} 
Most SRVs are found in the period range from 30 to about 
150\,days. However, there are a few exceptions that pulsate with, in most 
cases, a rather low light amplitude (i.e.~below the 2.5\,mag limit) but 
untypically long periods of 300 to 400\,days. This group must be clearly 
distinguished from SRVs with long secondary periods (see below).

\item {\it SRVs with amplitudes exceeding the 2.5 mag limit in their visual 
light change.} 

According to the GCVS definition, these stars should be 
miras, not SRVs.  However, the GCVS includes 57 SRVs with a listed amplitude exceeding 
2.5\,mag (in the blue or visual range). We note that the GCVS lists the maximum amplitude 
found in the literature and this is not necessarily the typical amplitude. 
It is possible that these stars have variable amplitudes.

\item {\it Miras with variable periods.} 
There are some miras that systematically change their period, either towards 
shorter or longer periods. This is normally interpreted as an indication 
of a recent He-shell flash (e.g.\,Wood \& Zarro 
\cite{WZ81}).  AAVSO light curves spanning several decades have helped 
to identify a few miras probably falling into this group 
(e.g.\,Hawkins et al.\,\cite{HMF01}).

\item {\it Miras with secondary maxima.} 
A small number of miras show a secondary maximum as a regular
part of their light curves.  Keenan et al.\,
(\cite{KGD74}) argued that the pulsation period which would put these stars
on the normal spectral type vs. period relation for miras is actually
half the period between the deep minima i.e. they there is only one maximum
per normal period. A similar result was found by Feast et al. 
(\cite{FRC82}) from the infrared color-period relation.

We note that a common feature of the light curves of many mira
variables is a small and irregular bump on the rising part of the light
curve (e.g. Lockwood \& Wing \cite{lw}).  We do not include the miras
with small bumps in the present category.

\item {\it SRVs with long secondary periods.} 
About 25\% of SRVs exhibit a long secondary period in their light curves,
the secondary period being typically a factor 10 longer than the primary 
period (Wood et al.\,\cite{WOK04}). A catalogue of 
solar vicinity SRVs with large amplitude secondary periods was presented 
by Houk (\cite{Houk63}). 
The MACHO survey of the Large Magellanic Clouds (LMC) and similar 
programs recently brought this group into focus, when it was discovered 
that the secondary periods follow a period-luminosity relation (sequence D 
of Wood et al.\,\cite{Wood99}).  These stars have been recently discussed in detail by Wood et 
al.\,(\cite{WOK04}), but the origin of the secondary period remains
a mystery.  

\end{itemize}

\begin{table}
\caption{Sample description. Column 2 lists the reason for selecting this 
star: LP SRV = SRV with period $\ge$300 d; LA SRV = SRV with amplitude 
above 2.5 mag; var. period = mira with variable period; s max = mira 
with a secondary maximum in the light curve; LSP = variable with long secondary period. 
Column 3 gives the variability class according to the GCVS. Periods and 
spectral types are from the GCVS except for W Nor.} \label{t:sample}
\begin{tabular}{lcccc}
Name        & anomaly     & Classif. & Period [d] & Sp. Type\\
\hline 
\noalign{\smallskip}
R Dor       & LP SRV      & SRb & 338,175 & M8IIIe\\
VZ Vel      & LP SRV      & SRa & 317 & M6e \\
WW Cen      & LP SRV      & SRb & 304,150 & M5-M7 \\
W Hya       & LP SRV      & SRa & 361 & M7.5e-M9ep \\
"           & LA SRV      & & & \\
T Cen       & LA SRV      & SRa & 90 & K0:e-M4IIe \\
L$^{2}$ Pup & LA SRV      & SRb & 141 & M5IIIe-M6IIIe \\
R Cen       & var. period & M & 546 & M4e-M8IIe\\
"           & s max       & & & \\
R Nor       & s max       & M & 507 & M3e-M6II \\
W Nor       & LSP         & SRb & 135,1300$^{1}$ & M4/5(III)\\
R Car       & --          & M & 309 & M4e-M8e \\
S Car       & --          & M & 149 & K5e-M6e \\
RR Sco      & --          & M & 281 & M6II/IIIe-M9 \\
R Hor       & --          & M & 408 & M5e-M8eII/III \\
\noalign{\smallskip}
\hline 
\end{tabular}
$^{1}$ Periods from Olivier \& Wood (\cite{OW03}).
\end{table}

In this paper, we investigate 9 LPVs that show -- according to the literature -- 
one of the behaviors listed above in 
their light variations. Their 'anomalies' together with some basic stellar 
data are listed in Table \,\ref{t:sample}.  Additionally, we present 
results on four rather typical, bright, southern miras observed 
in the course of this project. The stars selected also allow us to extend beyond 500 days 
the period range of miras whose velocity variations have been investigated.  A 
detailed description of each star's behavior, as found in the 
literature, will be given in Section~\ref{results_section}.

\section{Observations and Data Reduction}

Time series of infrared spectra in the H band were obtained in 2001 and 
2002 with the 74inch telescope at Mount Stromlo Observatory, Australia. 
The NICMASS detector, successfully used for a preceding program at Kitt 
Peak National Observatory (Joyce et al. \cite{Joyce98}), was used at 
the coud\'e focus of the telescope.  The resolution was set to 37000.  
The standard infrared observation technique was used. Spectra of each 
star were obtained at two different slit positions to allow sky 
subtraction. The spectra covered the range between 16280 and 16330\,{\AA}, 
including a number of second overtone CO lines, some OH lines, and a few 
metallic lines. This program used exactly the same instrumental setup 
as our monitoring program on 47\,Tuc variables (Lebzelter et al.\,
\cite{LWHJF04}). We refer to that paper for a further description of 
the observations and sample spectra. Our monitoring program came to an 
unexpected early end when Mount Stromlo Observatory was destroyed by 
a bushfire in early 2003. One additional spectrum for R Hor was
observed later in 2003 with the Phoenix spectrograph on Gemini South
using a similar wavelength range and a slightly higher resolution.

The bright stars $\alpha$ Cet and $\delta$ Oph have been used as primary 
velocity standards (Udry et al.~\cite{Udry99}). Velocities of the 
variables were determined by a cross correlation technique, using the 
IRAF task fxcor. Typical velocity uncertainties, determined from multiple 
observations of some stars in the same or consecutive nights, were found 
to be $\sim$0.4 kms$^{-1}$.

Light curve data have been taken from the data bases of the AAVSO (http://www.aavso.org),
the AFOEV (http://cdsweb.u-strasbg.fr/afoev) and the ASAS project (http://www.astrouw.edu.pl/gp/asas/asas.html). 
Phases were calculated with the periods listed in Table \ref{t:sample}, 
mostly from the GCVS.

\section{Velocity and light curves}\label{results_section}

Velocity curves of all 13 sample stars can been seen in Figs.\,\ref{rdor} 
to \ref{rhor}.   For each star, we briefly review the properties
that are relevant to the light and velocity change, as found in the 
literature.  We discuss the velocity curve of each object. We 
searched for line doubling, detected in several miras before (e.g. 
Hinkle et al.\,\cite{HHR82}), and detected it in a few stars of our 
sample.  However, we note that the velocity resolution plus 
the crowding and rather small wavelength range used in our observing 
program would not have allowed us to detect line doubling in every case. 

\subsection{R Dor}
At a first glance, the largest star in the sky (Bedding et al.\, 
\cite{Bedding97}) seems to be much more like a mira than a SRV. Its period 
of 338\,d falls within the mira regime, and its late spectral type is 
also more typical of miras than SRVs (Feast \cite{Feast96}). The star's 
spectrum sometimes shows emission lines. The GCVS lists a spectral type 
of M8IIIe, while Crowe \& Garrison (\cite{CG88}) report a spectral class 
of M8 but no emission lines.  Blum et al. (\cite{Blum03}) noted that 
this object, although one of the coolest AGB stars not classified as a 
mira, shows no evidence of H$_{2}$O absorption in its spectrum.

R Dor's light amplitude is smaller than that of typical miras, and its 
light curve shows phases of significant irregularity.  Bedding et al.\,
(\cite{Bedding97}) suggested that R\,Dor is actually near the edge of the 
Mira instability strip.  The star seems to switch its dominant pulsation 
mode within a few cycles, showing periods of 332 and 175\,d (Bedding et 
al.\,\cite{Bedding98}), which were attributed by Bedding et al.~to the
first and third overtones. However, with its 332\,d period the 
star nicely falls onto the PL-relation for miras in the LMC 
(Wood \cite{Wood00}), which is interpreted as fundamental mode pulsation
(see Section~\ref{pl-rel}).  
 
Olofsson et al. (\cite{Olof02}) analyzed the CO radio line profiles of 
various transitions. They found it very difficult to fit the profile of 
this star with their standard model. One of the explanations they 
suggested was that the star has highly variable mass loss. 

\begin{figure}
\centering
\resizebox{\hsize}{!}{\includegraphics{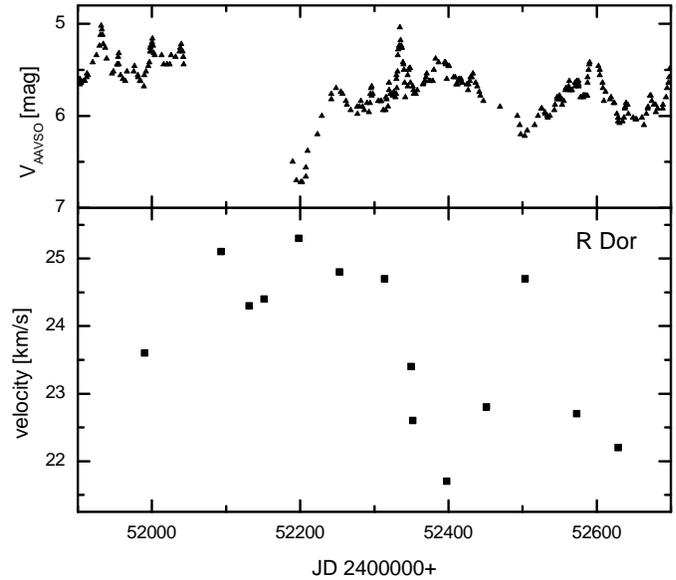}}
\caption{Velocity (bottom) and light (top) variations of R Dor. Light 
curve data were kindly provided by the AAVSO and have been smoothed to
10 day means.}
\label{rdor}
\end{figure}

Similarly, the light variability seen at the time of our velocity monitoring 
does not have a single period (Mattei \cite{Mattei04}, see Fig.\,\ref{rdor}, 
upper panel).  The velocity change appears to reflect the 
light change after JD2452300, a behavior typically observed in semiregular variables 
(Lebzelter et al.\,\cite{LKH00}). Interestingly, the deep minimum in 
the light change around JD 2452200 is not that obvious in the velocity 
change. This feature has also been found in a few SRVs before (Lebzelter 
et al.\,\cite{LKH00}).

\subsection{VZ Vel}
\begin{figure}
\centering
\resizebox{\hsize}{!}{\includegraphics{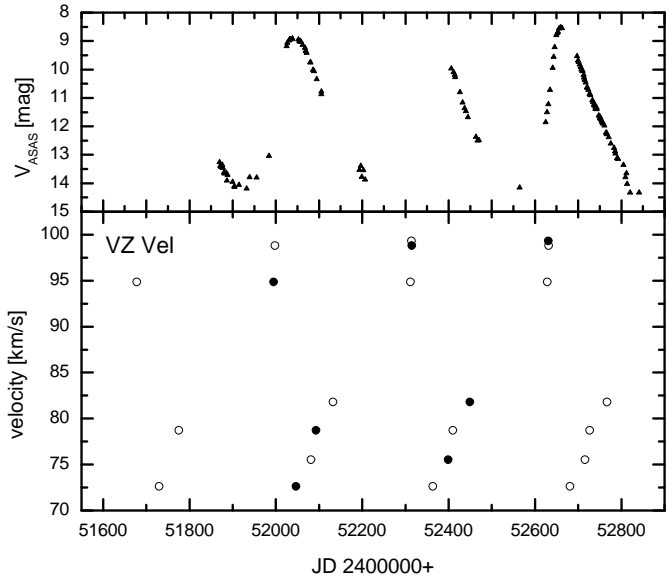}}
\caption{The light and velocity variations of VZ Vel. 
Upper panel: Light change from the ASAS data base. Lower panel:
Solid symbols show velocities
at their date of measurement while open symbols show the points shifted
by an integral number of 317d periods.}
\label{vzvel}
\end{figure}

VZ~Vel is classified as a SRV.  However, like R\,Dor, 
the long 317\,d period of VZ~Vel, its emission line 
spectrum, and its IRAS LRS 
class make it look more like a mira than a SRV 
(Kerschbaum \& Hron \cite{KH96}). We assume that the classification as SRV in the 
GCVS is based on an early measurement of its light amplitude, which is below 2.5 mag (Payne 
\cite{Payne28}). Opposite to this, more recent observations from the
ASAS project (Pojmanski \cite{Pojmanski02}) show a much larger light
amplitude and a light change similar to a mira (see Fig.\,\ref{vzvel}, upper
panel).

The velocity curve of VZ~Vel, which covers two cycles, is shown in Fig.\,\ref{vzvel},
lower panel.  
The shape and amplitude of the velocity curve clearly 
favour the star's classification as a mira.  Line doubling was not detected.
Based on the ASAS light curve and our velocity curve it is very obvious
that VZ Vel has to be classified as a mira today. It is not clear why
the star showed such a small amplitude in the observations made by
Payne (\cite{Payne28}).
  
\subsection{WW Cen}
The GCVS lists a single 304\,d period for WW Cen.  However, Gaposchkin 
(\cite{Gapo52}) detected secondary maxima in the light curve and
suggested that the star had a secondary period of about 150 days. Although
AAVSO measurements do not provide any definite conclusions about the 
star's period, recent measurements may support the idea of a second period. 
While the spectral class in the GCVS mentions no emission, Bidelman \& 
MacConnell (\cite{BM82}) detected emission lines in the spectrum of this star.

\begin{figure}
\centering
\resizebox{\hsize}{!}{\includegraphics{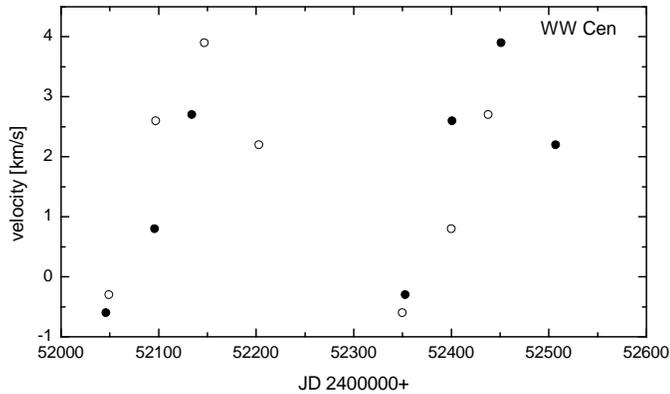}}
\caption{The velocity curve of WW\,Cen.  Symbols are as in Fig.\,\ref{vzvel},
using a period of 304 days.}
\label{wwcen}
\end{figure}

The velocity curve for WW Cen (Fig.\,\ref{wwcen}) clearly shows a period 
close to 304 days, although a period of 150 days can not be entirely
ruled out because of the incomplete phase coverage. 
The velocity amplitude is about 5\,kms$^{-1}$ and the curve is most 
probably sinusoidal, both these features being typical of semiregular variables. 

\subsection{W Hya}
\begin{figure}
\centering
\resizebox{\hsize}{!}{\includegraphics{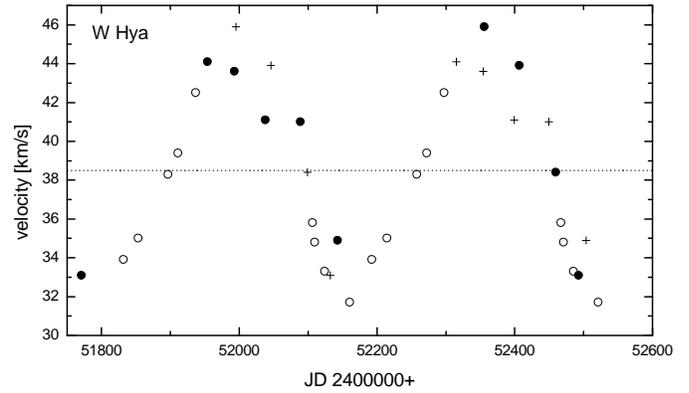}}
\caption{Velocity variations of W\,Hya.  Filled circles are new data from Mount Stromlo
Observatory while plus signs show these measurements shifted by an integral number of 361 day periods. 
Open symbols mark Fourier transform spectroscopy data from Kitt Peak National Observatory 
obtained in 1984/85, but shifted to the current dates.   The 
dotted line indicates the center of mass velocity.}
\label{whya}
\end{figure}

When compared to other SRVs, both the period (361\,d) and amplitude (3.9 
mag in $V$ according to GCVS) of W Hya, make this star outstanding. 
The pulsational behavior of W\,Hya has been studied by Hinkle et 
al.\,(\cite{HLS97}), and we refer to that paper for a detailed description 
of W\,Hya's stellar parameters. We acquired a second velocity time series 
of this star since the one obtained in the mid-80s sampled only a limited 
portion of the velocity cycle.  A complete velocity curve obtained 
by combining the data sets is shown in Fig.\,\ref{whya} and the 361 day
period is clearly evident.

The velocity curve of W\,Hya is rather sinusoidal with 
an amplitude of about 15\,kms$^{-1}$.  The two parts of the curve, 
obtained 17 years apart, fit well together, showing a high degree of 
periodicity in the variation of this star. Scatter along the velocity 
curve reflects small irregularities in the light curve.  The light curve 
parameters of this star (long period, large amplitude, periodic variations) 
closely resemble those of a mira.  However, the semiregular classification
seems more appropriate because the amplitude and sinusoidal shape 
of the velocity curve are clearly distinct from those of a mira.

\subsection{T Cen}
In addition to the large $V$ amplitude of 3.5 mag, the semiregular variable 
T\,Cen has a remarkable 
early spectral type during maximum (K0). From near-infrared colors, Lancon 
\& Mouhacine (\cite{LM02}) derived effective temperatures ranging between 
3600 and 4000\,K.  The star has sometimes been classified as a mira 
(e.g.\,Bidelman \& Ratcliffe \cite{BR54}). As in most miras, emission 
lines of hydrogen appear regularly in the star's spectrum (e.g.\,Keenan \& 
Landi Dessy \cite{KD66}). The 90\,d period seems to be very stable (Lacy 
\cite{Lacy73}, Kiss et al.\,\cite{KSCM99}).

\begin{figure}
\centering
\resizebox{\hsize}{!}{\includegraphics{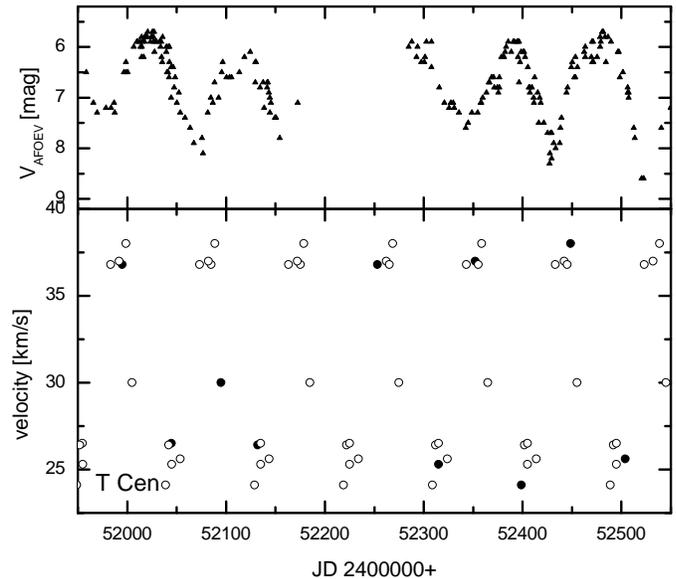}}
\caption{Velocity variations of T Cen.  Symbols are as in Fig.\,\ref{vzvel},
using a period of 90 days.}
\label{tcen}
\end{figure}

Because of the 90\,d period of the star, our velocity curve of T\,Cen 
(Fig.\,\ref{tcen}) consists of data from six cycles.
Obviously the repeatability of the variation is very high. Unfortunately, 
the data are concentrated near maximum and minimum velocity so that
a definite conclusion 
about the shape of the velocity curve cannot be drawn.  
The amplitude is about 14--16 kms$^{-1}$. No indication of line 
doubling was found.

\subsection{L$^{2}$ Pup}
\begin{figure}
\centering
\resizebox{\hsize}{!}{\includegraphics{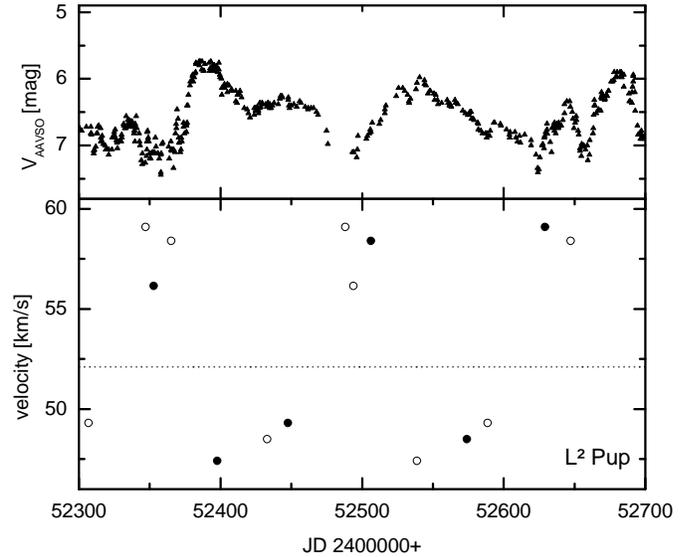}}
\caption{Velocity and light variations of L$^{2}$\,Pup. 
Symbols are as in Fig.\,\ref{vzvel},
using a period of 141 days.
The dotted line in the lower panel marks the center of mass velocity 
from circumstellar radio emission. Light curve data were kindly
provided by the AAVSO.}
\label{l2pup}
\end{figure}

This star is the third SRV in our sample with an amplitude exceeding 
the 2.5\,mag mira limit. The GCVS gives a $V$ amplitude of 3.6 mag and
a period of 141\,d.  From 
the recently obtained AAVSO light curve (Mattei \cite{Mattei04}), presented 
in the upper panel of Fig.\,\ref{l2pup}, one would estimate a somewhat 
smaller light amplitude of about 2.5\,mag. A light curve presented by 
Bedding et al.\,(\cite{Bedding02}) shows that over more than 10$^{4}$ 
days the amplitude never exceeded 2.5\,mag. However, Bedding et 
al.\,(\cite{Bedding02}) also note that the mean visual brightness of this star 
has faded significantly since about 1994. As the period of L$^{2}$ Pup 
remained rather constant over that time (and probably even since its 
discovery in 1872), Bedding et al.~conclude that the dimming is due to 
obscuration by dust.  Obviously, the star has still not recovered from 
this dimming.

Jura et al.\,(\cite{JCP02}) obtained mid infrared images of this star and 
detected an extended, asymmetric morphology.  To explain mass loss, time 
variations of the infrared flux, and time variations of the position 
angle of the optical polarization, Jura et al. proposed that L$^{2}$\,Pup 
is also pulsating in a nonradial mode.

Only six velocity measurements were obtained for this star. The 
data mostly sample phases of maximum and minimum as shown in 
Fig.\,\ref{l2pup}. Our radial velocity data are consistent with a
141\,d period and a velocity amplitude of 12\,kms$^{-1}$, the latter being more typical
for a semiregular variable than for a mira.

\subsection{R Cen}
The light variations of this mira are very remarkable. Its light curve is 
double peaked with a main period of 546\,d and a secondary period of 
274\,d (e.g. Feast et al.\,\cite{FRC82}).  In the past 50 years the long 
period has been steadily getting shorter and is now around 505-510 days (Hawkins 
et al.\,\cite{HMF01}). This period change of about 1\,day\,yr$^{-1}$ has 
been interpreted as the result of  a recent He shell flash (see Wood \& 
Zarro \cite{WZ81}). Not only the period but also the amplitude of the 
$\approx$500\,d variation decreased in recent decades from more than 5 mag 
to currently about 2.5 mag (Hawkins et al.,\,\cite{HMF01}). 

\begin{figure}
\centering
\resizebox{\hsize}{!}{\includegraphics{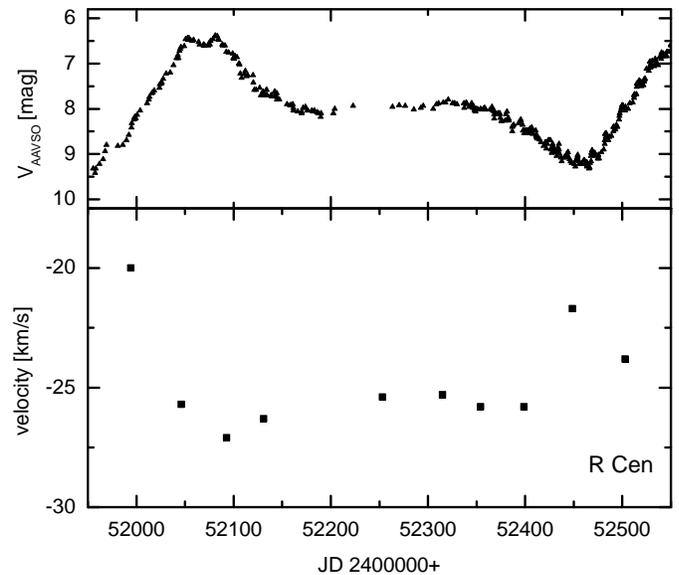}}
\caption{Velocity (bottom) and light (top) variations of R Cen. Light curve 
data (smoothed to 10 day means), which show a period of about 500\,d, were kindly provided by the 
AAVSO. }
\label{rcen}
\end{figure}

This drastic change is also reflected in the velocity variations of the 
star.  Untypically for a mira, the velocity amplitude is only about 
8\,kms$^{-1}$ i.e. the velocity curve of R\,Cen is more typical of a 
semiregular variable rather than a mira.  The
velocity variations mirror the light curve as shown in Fig.\,\ref{rcen}.
The long interval of almost no velocity variation, with a velocity spike
just before rising light, is very unusual.  Note that this velocity spike
corresponds to sudden infall to the star and is in the opposite sense
to the discontinuities in mira velocity curves which are caused by
a shock wave pushing matter outward.  We found no evidence for a shock wave
in the form of line doubling in R\,Cen.
There is clearly an unusual dynamical variation associated
with pulsation of this star, and there is some slight evidence for a
hump in the velocity curve as well as in the light curve.

While Keenan \& Landi Dessy (\cite{KD66}) report emission lines during all maxima, data obtained later by
Crowe (\cite{Crowe82}) show that H emission only occurs during the maxima corresponding to rising light i.e.
at intervals of $\sim$546\,d.
This is consistent with the fact that the velocity curves show emergence of
a velocity pulse at this time.

\subsection{R Nor}
\begin{figure}
\centering
\resizebox{\hsize}{!}{\includegraphics{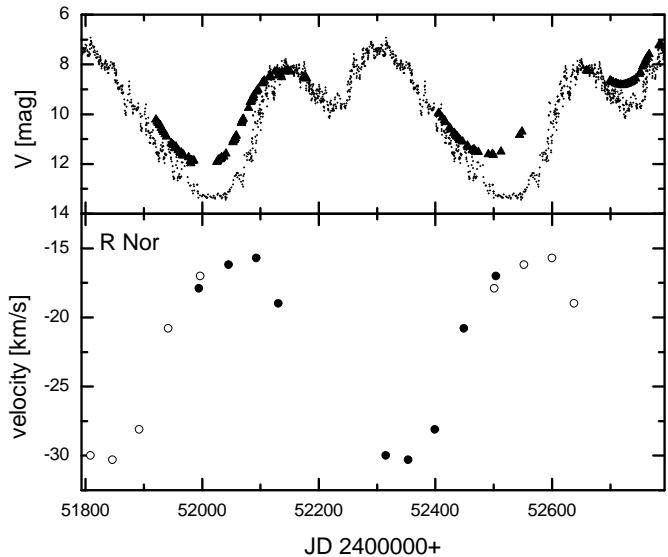}}
\caption{Top panel: The light variability of R Nor. Triangles indicate
the parallel measurements from the ASAS database. Overlayed is
a mean AAVSO light curve (small crosses) constructed from earlier data 
as there are unfortunately no parallel AAVSO data. 
This mean light curve helps to illustrate the star's typical light change.
Bottom panel: Velocity curve 
of R Nor.  Solid circles show observations plotted
on the date measured while open circles show the measurements shifted by an integral
number of 507\,d periods.}
\label{rnor}
\end{figure}

Among the stars in our sample, R\,Nor shows the most remarkable light curve
humps. Unfortunately, there is no complete light curve
coverage during the time of our observations. 
In the uppermost panel of Fig.\,\ref{rnor} we thus combine parallel photometric
observations from the ASAS database with a representative 
part of the AAVSO light curve of R Nor to illustrate its outstanding 
variability. Secondary maxima are clearly visible during most cycles. 
The current data fit well with the average light curve except for somewhat
less expressed minima in the ASAS data. This may result from combining
CCD (ASAS) and visual (AAVSO) brightness measurements.

The star is a visual binary (GCVS, Proust et al.\,\cite{proust81}), a fact
recently confirmed by an improved analysis of the Hipparcos data (Pourbaix 
et al.\,\cite{pourbaix}).  The companion has a $V$ magnitude of 13.8. The 
$V$ brightness of the mira varies between 7.7 and 11.55\,mag, the secondary 
minimum is $V$$=$9.5\,mag (Celis \cite{celis77}). Some period variability 
was noted by Templeton \& Mattei (\cite{TM02}).

The velocity curve of R\,Nor is shown in Fig.~\ref{rnor}.
Although it has been suggested that the ``normal'' period for miras like R\,Nor and R\,Cen
is the semi-period (i.e. 253.5\,d for R\,Nor - see Sect.~\ref{sect1} and Jura \cite{jura94}), 
it is clear from the velocity curve of R\,Nor (and R\,Cen) that the true pulsation period of these 
stars is the full period i.e. the  interval between alternate maxima or minima (507\,d for R\,Nor).

No clear case of line doubling was detected in the spectra or in the
cross-correlation profiles of R\,Nor.  However, for spectra between JD2452040
and JD2452130, the cross correlation profile is clearly broader than at the
phases before.  The velocity is beginning to change rapidly at these phases as
the velocity infall is reversed (possibly due to an emerging shock front) and
the broad lines suggest that there is a velocity gradient through the
line-forming region of the atmosphere.

\subsection{W Nor}
The variability of W Nor was discovered about a century ago (Pickering et 
al.\cite{picke01}), but only a few investigations of this star are found 
in the literature.  Most 
remarkable is the star's long secondary period of about 1300 days,  
approximately ten times its primary period of 135\,d (Houk \cite{Houk63}; 
Olivier \& Wood \cite{OW03}).  Accordingly, we have observed this star 
as a representative of SRVs with long secondary periods.
Circumstellar material is obviously present, since 
silicate emission was reported by Sloan \& Price (\cite{sp98}).  

\begin{figure}
\centering
\resizebox{\hsize}{!}{\includegraphics{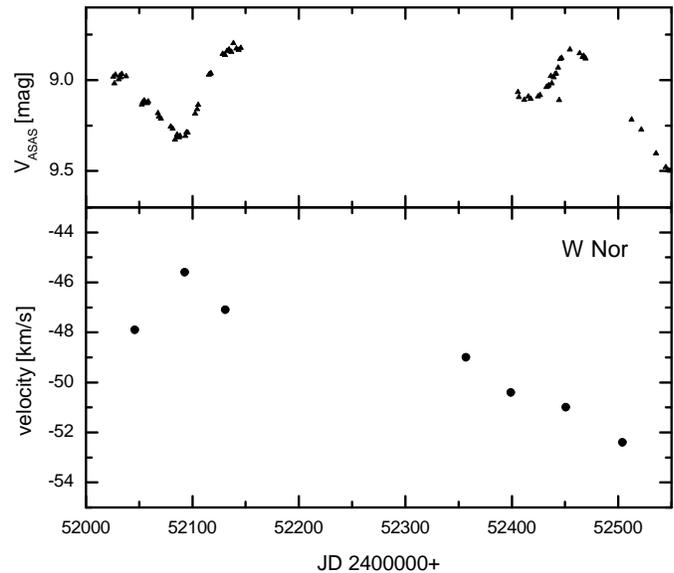}}
\caption{Light (upper panel) and velocity (lower panel) variations of W\,Nor.
Light curve data are from the ASAS database.}
\label{wnor}
\end{figure}

The velocity curve of W\,Nor (Fig.\,\ref{wnor}) is clearly dominated by a
variation on a time scale longer than the 135\,d primary period, but our
monitoring covers only about one-third of the secondary 1300\,d light-variation
period.  Variations on the shorter time scale are perhaps seen on top of the
long-term changes. The velocity amplitude shown in Fig.\,\ref{wnor} is at least
7\,kms$^{-1}$, and that variation seems to arise mostly from the 1300\,d
period.

Although the plotted light change does not indicate any long period variation
it can be seen very well on the whole ASAS dataset. Our measurements seem to be
close to the light maximum of the longer period, the next minimum is reached
around 2452800 (compare ASAS database: http://www.astrouw.edu.pl/~gp/asas/asas.html).

\subsection{R Car}

R\,Car is a typical mira with a slight bump on the rising part of the light
curve, typical of many miras.  The bump is more distinct at 4.9\,$\mu$m than in
the visual (Smith et al.\,\cite{SLCL02}). In current AAVSO data (Mattei
\cite{Mattei04}) the hump is barely visible.  The period seems to be well
determined and stable (Mennessier et al.\,\cite{MBM97}; Mattei
\cite{Mattei04}).  The star is a visual binary (Proust et
al.\,\cite{proust81}).

\begin{figure}
\centering
\resizebox{\hsize}{!}{\includegraphics{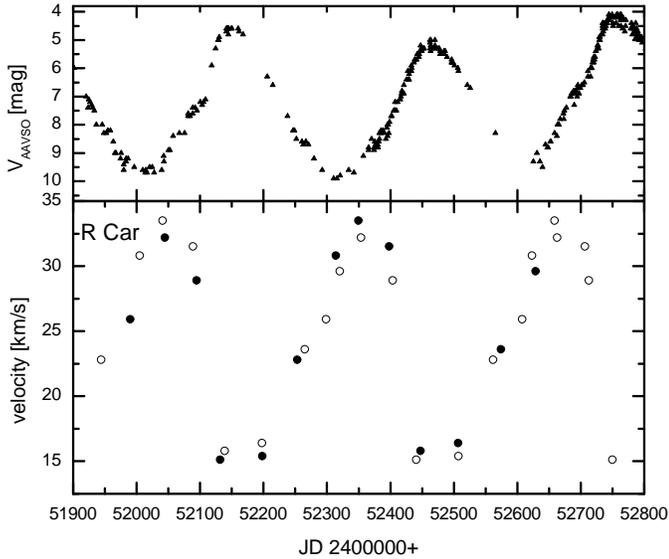}}
\caption{Velocity (bottom) and light (top) variations of R\,Car.
Symbols in the velocity plot are as in Fig.\,\ref{vzvel},
using a period of 309 days.}
\label{rcar}
\end{figure}

Fig.\,\ref{rcar} shows the velocity variations of this star.
The shape and amplitude are mira-like although a final decision 
between a continuous and a discontinuous velocity curve cannot be made 
with our data.  The velocity curve seems to show a scatter 
exceeding the typical uncertainty of
our measurements of 0.4\,kms$^{-1}$.  Examination of the velocity curve shows that the
scatter is due mostly to cycle-to-cycle variations.
Similar scatter has been noticed previously in the velocity curves of 
several miras (e.g.\,Hinkle et al.\,\cite{HLS97}).  
We found no indication of line doubling in our spectra.

\subsection{S Car}
S\,Car belongs to the group of short period miras with intermediate 
metallicity (Hron \cite{hron91}).  In agreement with the star's metal-poor 
nature, its space velocity is very large, namely about 288\,kms$^{-1}$ 
(e.g.\,Wallerstein \& Dominy \cite{wd88}).

Shinkawa (\cite{shinkawa73}) obtained a complete velocity curve for S\,Car
from atomic lines around 8000\,{\AA}. The early spectral type
of this star and its reduced metallicity relative to the solar value
allowed her to investigate a number of almost unblended lines
in that spectral region. For a few lines, two components could be resolved.
Doubling of these lines in S\,Car was also found by Gillet et 
al.\,(\cite{gillet85}), but was interpreted by these authors as a single 
absorption line with an emission on top of it.

\begin{figure}
\centering
\resizebox{\hsize}{!}{\includegraphics{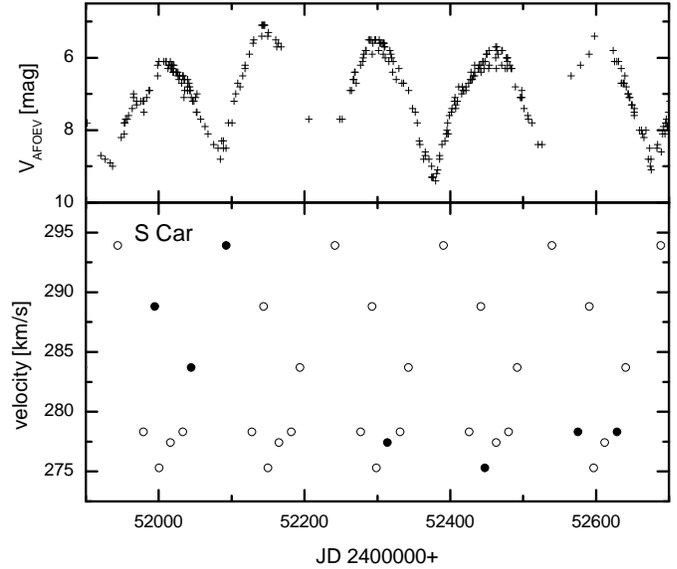}}
\caption{Velocity (bottom) and light (top) variations of S\,Car.
Symbols in the velocity plot are as in Fig.\,\ref{vzvel},
using a period of 149 days.}
\label{scar}
\end{figure}

Our velocity curve of this star is presented in Fig.\,\ref{scar}. Velocity
amplitude is about 20\,kms$^{-1}$, with a clear line doubling observed in
at least one spectrum. The curve presented by Shinkawa (\cite{shinkawa73})
has the same shape and also a similar amplitude, but significantly larger error bars, 
so that a more detailed comparison of the
velocity data, e.g.~concerning a possible phase shift, is not possible.

Velocities of \ion{Fe}{ii} lines in the ultraviolet were measured by 
Wood \& Karovska (\cite{wk}). Between phase 0 and 0.4 the centroid 
velocity of the line at 2625.7\,{\AA} varied by about 20 to 25\,kms$^{-1}$, 
an amplitude similar to the near-infrared CO lines, but no clear shape 
of the velocity curve is visible in their data.

\subsection{RR Sco}

\begin{figure}
\centering
\resizebox{\hsize}{!}{\includegraphics{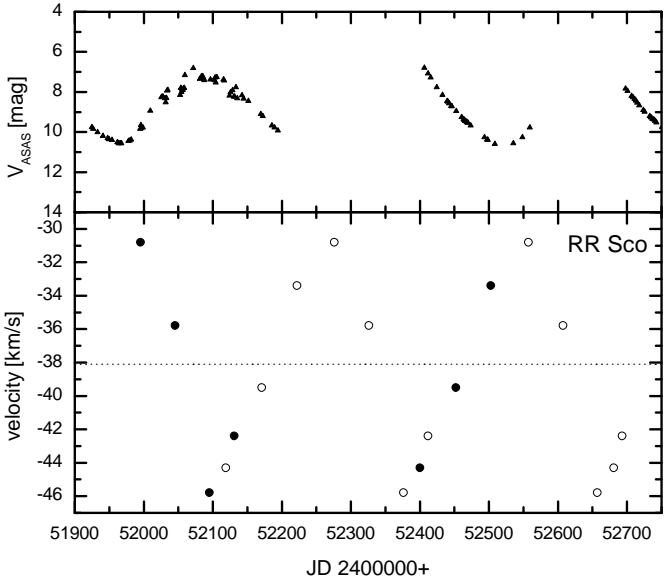}}
\caption{Velocity (bottom) and light (top) variations of RR\,Sco.
Symbols in the velocity plot are as in Fig.\,\ref{vzvel},
using a period of 281 days.  The dotted line 
marks the center of mass velocity determined from circumstellar radio emission. 
Light curve data are from the ASAS database.}
\label{rrsco}
\end{figure}

RR\,Sco is a typical mira with no obvious anomalies. The light curve,
with a period of 281\,d, is symmetric (Vardya \cite{vardya88}).  
From its velocity curve (Fig.\,\ref{rrsco}) RR\,Sco looks like a typical 
mira except for its relatively small amplitude of $\approx$15.5\,kms$^{-1}$. 
However, the amplitude may be slightly larger because a data point at 
maximum light is missing. The correlation peak for the spectrum near JD2452000
(just before velocity velocity turnaround) 
is asymmetric which may suggest line doubling that was unresolved. 
This again would favour a slightly larger amplitude for the velocity curve. 
As for most miras (e.g.\,Lebzelter \& Hinkle \cite{LH02}), the velocity 
curve crosses the center of mass velocity (taken from Young \cite{young95}) 
around phase 0.4.

\subsection{R Hor}

R\,Hor is one of the brightest southern miras. 
Percy \& Colivas (\cite{PC99}) searched for period changes in this star, 
but found only a rather large point-to-point scatter in the O-C diagram.

\begin{figure}
\centering
\resizebox{\hsize}{!}{\includegraphics{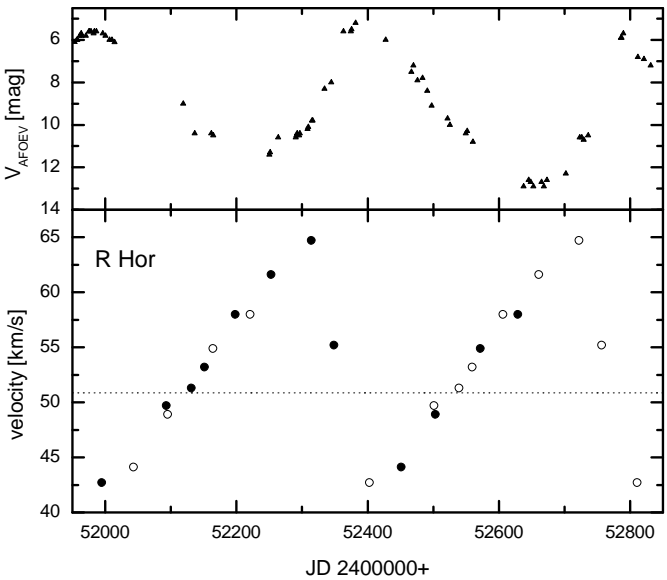}}
\caption{The same as Fig.~\ref{rcar} but for R\,Hor with a period of
408\,d.Velocity phase curve of R\,Hor with a period of 408\,d.  The dotted 
line marks the center of mass velocity from circumstellar radio emission 
(see text).}
\label{rhor}
\end{figure}

The velocity curve in Fig.\,\ref{rhor} indicates that this star is a typical 
mira. Although no clear case of line doubling was detected, the 
cross-correlation peak for the velocity near JD2452340 is asymmetric and 
likely indicates unresolved line doubling.  The velocity amplitude of 23\,kms$^{-1}$  
is comparable to most miras.  A center of mass velocity 
was taken from radio CO(3-2) observations presented by Young (\cite{young95}). 

To finish this section, we show in Table \ref{t:results} a summary of the results 
from the velocity curves.

\begin{table}
\caption{Characteristics of the velocity curves of the sample stars. 
Column 3 gives the total velocity amplitude measured in kms$^{-1}$. 
Column 4 indicates whether line doubling was observed. If a star
was not observed at the expected phase of line doubling, we state "not 
observed". The last column describes the shape of the velocity curve} 
\label{t:results}
\begin{tabular}{lcccc}
Name & Classif. & velocity & line & Shape \\
 & GCVS & amplitude & doubling & vel.-curve\\
\hline 
\noalign{\smallskip}
R Dor & SRb & 4 & no & continuous\\
VZ Vel & SRa$^{1}$ & 22 & no & prob.\,discont.\\
WW Cen & SRb & 5 & no & prob.\,cont.\\
W Hya & SRa & 14.5 & no & continuous\\
T Cen & SRa & 14 & not observed & prob.\,cont.\\
L$^{2}$ Pup & SRb & 12 & no & continuous\\
R Cen & M & 8 & no & continuous\\
R Car & M & 18.5 & not observed & prob.\,discont.\\
R Nor & M & 15 & likely & prob.\,discont.\\
W Nor & SRb & $>$7 & not observed & ?\\
S Car & M & 19 & yes & discontinuous\\
RR Sco & M & 15.5 & likely & discontinuous\\
R Hor & M & 23 & likely & discontinuous\\
\noalign{\smallskip}
\hline 
\end{tabular}
$^{1}$ This star seems to be a mira (see text).
\end{table}

\section{The logP-M$_{K}$-diagram}\label{pl-rel}

\begin{figure*}
\centering
\resizebox{\hsize}{!}{\includegraphics{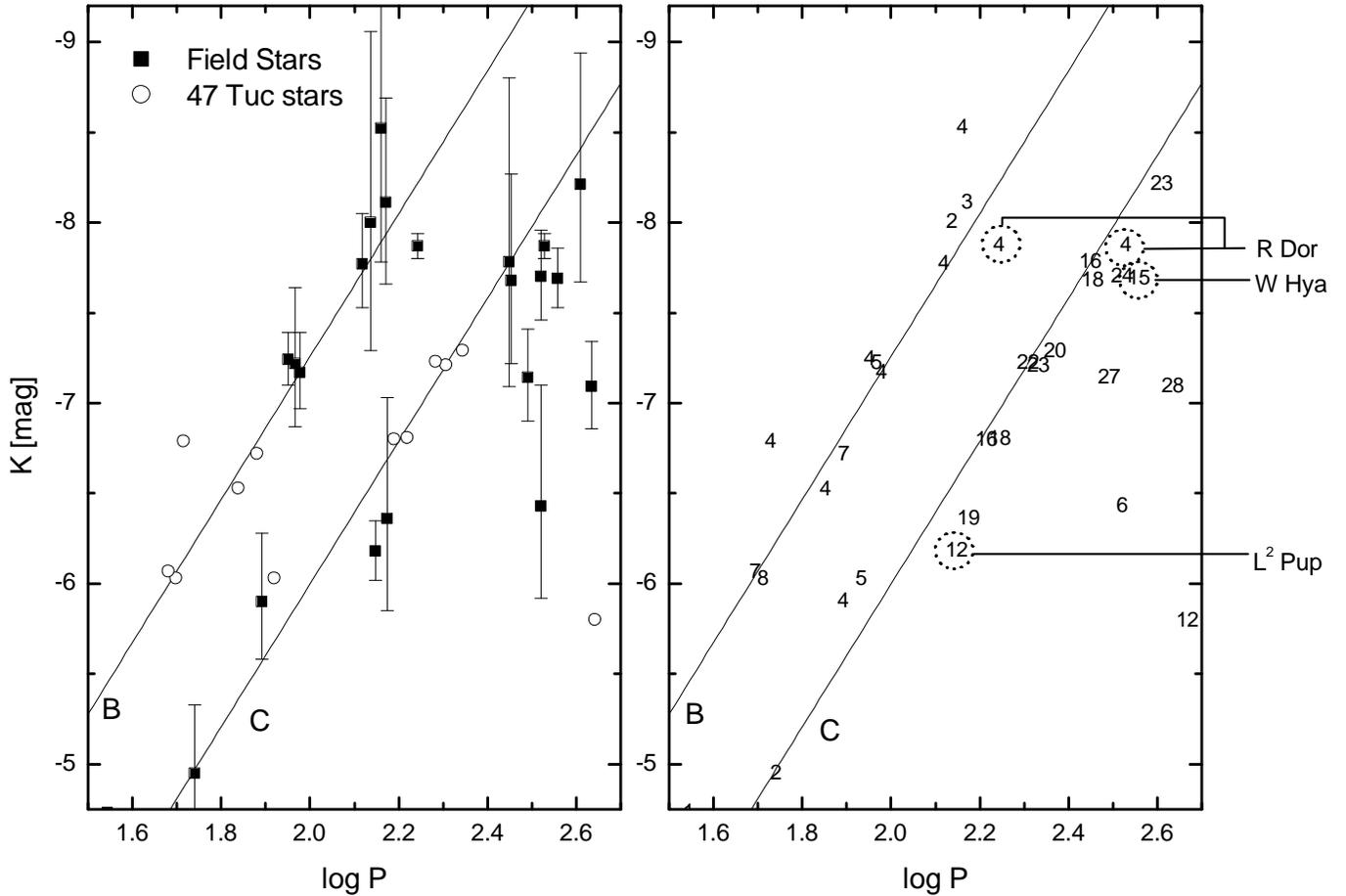}}
\caption{Period-luminosity diagram for 47 Tuc and field LPVs with known 
velocity variations.  Solid lines indicate the approximate location of 
the logP-K-sequences B and C found in the LMC (Wood \cite{Wood00}).  In 
the left panel, stars are divided by symbol into field and globular 
cluster LPVs. Error bars have been calculated from $\sigma$($\pi$) as 
given in the Hipparcos catalogue. 
The dash-dotted line marks the LMC mira relation derived by Whitelock 
\& Feast (\cite{WF00}). The right panel gives the velocity amplitude 
measured for each object. Some stars discussed in the text have 
been identified.}
\label{logpk}
\end{figure*}

To have an additional tool for the analysis of our results we determined 
absolute $K$ magnitudes for all stars with good Hipparcos parallax 
measurements and plotted them in a logP-$M_{K}$-diagram (left panel of 
Fig.\,\ref{logpk}). $K$ magnitudes were taken from 
Bagnulo (\cite{Bagnulo96}), Fouque et al. (\cite{Fouque92}),
Kerschbaum et al. (\cite{Kerschbaum01} and references therein)
and Whitelock et al. (\cite{Whitelock00}). Absolute $K$ magnitudes
were then calculated using Hipparcos parallaxes (Pourbaix et al.
\cite{pourbaix}). The selection criterion was that the parallax of the 
star must be larger than two times the parallax error.  Furthermore, we 
added all stars from our previous studies that fulfill this criterion. 
Finally, AGB variables from the globular cluster 47\,Tuc  (Lebzelter et 
al.\,\cite{LWHJF04}) also are plotted. In that paper, we presented 
for the first time the velocity amplitude as a function of the star's 
location in the logP-$M_{K}$-diagram, which, in the case of a globular 
cluster, is a function of the evolutionary path of an AGB star. 

The same was done in the right panel of Fig.\,\ref{logpk}, where the 
numbers in the figure correspond to the velocity amplitude of each star.  
All stars plotted in Fig.\,\ref{logpk} are listed in Tab.\,\ref{logpktab}.
Solid lines indicate the approximate location of the logP-$M_{K}$-sequences 
B and C found for LMC stars by Wood et al.~(\cite{Wood99}, \cite{Wood00};
see also Lebzelter et al.~\cite{LWHJF04}). These two sequences likely represent fundamental 
and first overtone mode pulsation. Sequence C is in good agreement with 
the mira logP-K-relation derived by Whitelock \& Feast (\cite{WF00}), 
combining the P-L-relation from the LMC with a zero point from the Hipparcos 
data (dash-dotted line).  We show in our paper on the 47\,Tuc variables 
(Lebzelter et al.\,\cite{LWHJF04}) that the velocity amplitude increases 
along sequence C, while stars on sequence B all have smaller amplitudes.

\begin{table}
\caption{Stars plotted in Fig.\,\ref{logpk}. Velocity amplitudes
(column 4) were rounded to the nearest integer. The last
column gives the reference to the velocity curve.} 
\label{logpktab}
\begin{tabular}{lcccc}
Name & log\,P & $M_{K}$ & vel.\,ampl. & ref. \\
 & & & [kms$^{-1}$] & \\
\hline 
\noalign{\smallskip}
R\,Aql & 2.454 & $-$7.68 & 18 & 2\\
RV\,Boo & 2.137 & $-$8.00 & 2 & 4\\
BC\,CMi & 1.544 & $-$4.73 & 1 & 4\\
S\,Car & 2.174 & $-$6.36 & 19 & 7\\
R\,Cas & 2.634 & $-$7.09 & 28 & 2\\
o\,Cet & 2.521 & $-$7.70 & 24 & 2\\
RS\,CrB & 2.520 & $-$6.43 & 6 & 5\\
W\,Cyg & 2.118 & $-$7.77 & 4 & 3\\
AF\,Cyg & 1.966 & $-$7.22 & 5 & 5\\
R\,Dor & 1.740 / 2.529 & $-$7.87 & 4 & 7\\
TX\,Dra & 1.892 & $-$5.90 & 4 & 4\\
g\,Her & 1.950 & $-$7.24 & 4 & 4\\
ST\,Her & 2.170 & $-$8.11 & 3 & 4\\
X\,Her & 1.978 & $-$7.17 & 4 & 4\\
R\,Hor & 2.610 & $-$8.21 & 23 & 7\\
W\,Hya & 2.558 & $-$7.69 & 15 & 7\\
R\,Leo & 2.491 & $-$7.14 & 27 & 1\\
SV\,Peg & 2.160 & $-$8.52 & 4 & 3\\
L$^{2}$\,Pup & 2.148 & $-$6.18 & 12 & 7\\
RR\,Sco & 2.449 & $-$7.78 & 16 & 7\\
ER\,Vir & 1.740 & $-$4.95 & 2 & 4\\
47\,Tuc V1 & 2.344 & $-$7.29 & 20 & 6\\
47\,Tuc V2 & 2.308 & $-$7.21 & 23 & 6\\
47\,Tuc V3 & 2.283 & $-$7.23 & 22 & 6\\
47\,Tuc V4 & 2.219 & $-$6.81 & 18 & 6\\
47\,Tuc V5 & 1.699 & $-$6.03 & 8 & 6\\
47\,Tuc V6 & 1.681 & $-$6.07 & 7 & 6\\
47\,Tuc V7 & 1.839 & $-$6.53 & 4 & 6\\
47\,Tuc V8 & 2.190 & $-$6.80 & 16 & 6\\
47\,Tuc V11 & 1.716 & $-$6.79 & 4 & 6\\
47\,Tuc V13 & 2.643 & $-$5.80 & 12 & 6\\
47\,Tuc V18 & 1.919 & $-$6.03 & 5 & 6\\
47\,Tuc V21 & 1.881 & $-$6.72 & 7 & 6\\
\noalign{\smallskip}
\hline 
\noalign{\smallskip}
\end{tabular}

\begin{small}
References: 1 - Hinkle \cite{Hinkle78}; 2 - Hinkle et al.\,\cite{HHR82}; 3 - Hinkle et al.\,\cite{HLS97};
4 - Lebzelter \cite{L99}; 5- Hinkle et al.\,\cite{HLJF02}; 6 - Lebzelter et al.\,\cite{LWHJF04};
7 - this paper.\\
\end{small}
\end{table}

We see that, with a few exceptions, our stars lie close to the two 
logP-$M_{K}$-relations.  The two sequences can also be nicely divided by 
the velocity amplitude found there: stars on sequence B (first overtone) 
show only small velocity amplitudes, in most cases around 4\,kms$^{-1}$. 
Results from field stars and variables in 47 Tuc are in general agreement.  

On the other hand almost all stars on sequence C (fundamental mode) show 
velocity amplitudes greater than 10\,kms$^{-1}$. An exception is R\,Dor 
(4\,kms$^{-1}$). Most of the field 
stars fall slightly below sequence C, with R\,Cas and R\,Leo being the 
most extreme cases.  There are two likely reasons for this 
difference. Firstly, the parallaxes have relatively large uncertainties, as can be seen 
from the error bars in Fig.\,\ref{logpk} (left panel). Note that the 
accuracy of Hipparcos parallaxes for AGB variables may suffer because in 
some cases the angular diameter is larger than their parallax (see a 
discussion in Whitelock \& Feast \cite{WF00}).  Secondly, in these cool 
stars the K band can be heavily affected by H$_{2}$O absorption and/or 
circumstellar dust absorption. Since the field stars on sequence B show no 
obvious offset, the last possibility could explain the offset from of 
stars from sequence C because they typically have cooler temperatures, 
more extended atmospheres, more mass loss and more circumstellar absorption.  
This may be particularly 
relevant for R\,Cas, which has a very late spectral type (M10) at minimum 
and strong water absorption (Aringer et al.\,\cite{AKJ02}).  In the 
following discussion, we assume that the stars with large velocity amplitudes
belong to sequence C even if they fall below this sequence in Fig.\,\ref{logpk}.

\section{Discussion}

\subsection{Long period SRVs}

This group of stars (R Dor, WW Cen and W Hya) all have
periods slightly longer than 300 days.  Only R Dor and W Hya have
parallaxes that allow them to be plotted in Fig.\,\ref{logpk}.  Both these
stars fall on the fundamental mode sequence, C.  Thus, we have here examples of
fundamental mode pulsators with smaller amplitudes of light and pulsation
velocity than miras with shorter periods lower down sequence C.  The reason for
the small amplitude at relatively high luminosities is unclear, but it may be
that these stars are more massive than the typical mira since a general feature
of pulsation models for LPVs is that increasing the mass tends to decrease the
instability (Fox \& Wood \cite{FW82}).  Although these stars have smaller
amplitudes than those of miras, all four show, or have shown at some time,
emission lines in their spectra, implying the presence of a shock in the stellar
atmosphere. Their amplitudes are therefore still quite large at times, indicating that the
modes involved are intrinsically unstable.  It is very unlikely that a
stochastic excitation mechanism (e.g. Dziembowski et al.~\cite{Dziem01}) 
could produce such large amplitudes in the fundamental mode.

R\,Dor has a second period which lies on sequence B. Its current small velocity
amplitude of only about 4\,kms$^{-1}$ is in good agreement with amplitudes
typically found in sequence B.  The relative amplitude of the two modes has
varied with time (Bedding et al.\cite{Bedding98}).  WW\,Cen also has a
secondary period which, like that of R\,Dor, is approximately half the length of
the primary period. 

VZ\,Vel currently has a velocity curve with a shape and
amplitude that identifies it as a Mira.  Its original classification as a SRV
suggests that the amplitude was much smaller in the past (with the same period
as at present). Currently one would classify this star as a mira, but
it may be related in some way to this group of objects.

It appears that a characteristic of these stars is that they
have multiple modes (fundamental and first overtone, sequences C and B,
respectively, being dominant), as well as time-variable amplitudes in a given
mode.  These stars may be in the evolutionary stage of switching from first
overtone to fundamental mode.  We note that the distribution of the 47\,Tuc
variables along the two pulsation sequences would suggest that mode switching
occurs only from first overtone to fundamental mode, and not the other way,
because the most luminous stars in 47\,Tuc are all found on sequence C.  The
period and luminosity at which mode switching occurs will increase with the
mass of the star.

However, for a number of stars mode switches in the opposite direction, i.e.
from fundamental to first overtone, are reported in the literature
(e.g.\,Kiss et al.\,\cite{Kiss00}). This may indicate that mode switches may
be not be driven by evolution alone. 

\subsection{Large amplitude SRVs}

We have analyzed three stars from this group: L$^{2}$\,Pup, 
T\,Cen and W\,Hya (which was also discussed in
the last section). According to the GCVS, these objects all 
have $V$ light amplitudes between 3.5 and 3.9\,mag. 
However, L$^{2}$ Pup, at least in recent times, has had a significantly smaller 
light amplitude. AAVSO data of W\,Hya and T\,Cen show that these two stars 
currently have light amplitudes that clearly exceed 3\,mag. All three stars
show similar velocity amplitudes of 14, 12 and 14\,kms$^{-1}$, respectively,
and clearly continuous, approximately sinusoidal velocity curves. 
Velocity curves of two additional large amplitude SRVs were presented in 
previous papers: X\,Oph (Hinkle et al.\,\cite{HSH84}) and SV\,Cas 
(Lebzelter et al.\,\cite{LKH00}). Both objects have velocity amplitudes in the range 
12$-$14\,kms$^{-1}$, and continuous rather than saw-tooth shape velocity curves.  

Two of the large amplitude SRVs in this study, L$^{2}$\,Pup and W Hya, are plotted in 
Fig.\,\ref{logpk}.  Both stars are found close to sequence C, although their 
luminosities are quite different. 
These large amplitude SRVs thus seem to be fundamental
mode pulsators, but their pulsation is not as powerful as in the miras so
that the velocity curves do not show the discontinuity associated
with the emergence of a powerful shock wave through the photosphere.

\begin{figure}
\centering
\resizebox{\hsize}{!}{\includegraphics{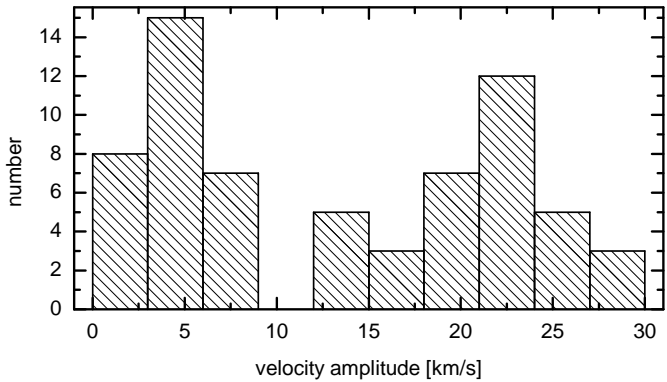}}
\caption{Number of LPVs versus near-infrared velocity amplitude.}
\label{velosdistr}
\end{figure}

In Fig.\,\ref{velosdistr} we show the distribution of near infrared velocity amplitudes for
all the LPVs we could find with measurements in the literature.  The ordinary
SRVs, which have amplitudes below 2.5 magnitudes and which make up all the
first overtone pulsators on sequence B and a few of the fundamental mode
pulsators at the bottom of sequence C, mostly have velocity amplitude below 10
kms$^{-1}$.  The miras lie in the peak with amplitudes around 23\,kms$^{-1}$,
and at higher amplitudes.  The large amplitude SRVs lie at the lower limit of
the mira range.

As with the long period SRVs discussed in the last section, there must be some
stellar parameter of these fundamental mode pulsators that causes them to have
smaller amplitudes than the fundamental mode miras.  We suggest that the reason
for the lower amplitude is a slightly higher mass for the fundamental mode SRVs
compared to the miras at a similar period or luminosity (see also the last
subsection).  Such a mass variation will cause the PL relations in Fig. 14 to
have a finite width, and should indeed be part of the reason for the observed
scatter.

Are the SRVs with light amplitudes exceeding 2.5\,mag. common?  There are 57
such SRVs (SRa or SRb) listed in the GCVS.  This is about 7\% of all M-type
SRVs (SRa and SRb only). Periods range between 90 and more than 400
days. However, their mean period is around 250\,days, thus closer to the miras
regime than the typical SRVs. We also compared IRAS colors of this group with
SRVs having amplitudes below 2.5\,mag, but found no significant difference.
Given the small number of stars in this group, it seems plausible that they are
of comparably high mass.

\subsection{Miras with secondary maxima}

Two stars in our sample belong to this group: R\,Cen and R\,Nor. 
R\,Cen currently has both a small light and velocity amplitude for a mira (only 
$\approx$ 2.5 mag. and 8\,kms$^{-1}$, respectively). The velocity curve (Fig.\,\ref{rcen}) also is
not typical of a mira, but rather mirrors the visual light changes, similar 
to the semiregular variables.  The light and
velocity amplitudes of R\,Nor are larger (7 mag and 15 \,kms$^{-1}$, respectively), 
with the velocity amplitude being at the lower end of the mira 
velocity amplitude distribution. 

Miras with double-peaked maxima are seen at high luminosities in the LMC, on
sequence C (Wood et al. \cite{Wood99}).  This suggests that the miras with
double-peaked maxima are of high mass (3--5 M$_{\odot}$).  The high mass would
also explain the relatively early spectral types (see Sect.~\ref{sect1}) since
the giant branch for massive stars is warmer than that for lower mass stars.
The bumps seen in the light curves of these stars are probably the analogs of
the bumps seen on bump Cepheid light curves, meaning that the bumps are the
result of a 2-to-1 resonance between the fundamental mode and the first
overtone.  Indeed, Fig. 4 of Wood et al. (\cite{Wood99}) shows that the long
period miras have a period ratio, fundamental to first overtone, of
approximately 2/1.

\subsection{Normal miras}

We reported observations of four
rather typical miras (S\,Car, R\,Car, RR\,Sco and R\,Hor). VZ\,Vel,
discussed above, also -- at least currently -- belongs to this group.
For RR\,Sco 
and R\,Hor we can compare the velocity curve with the center of mass 
velocity derived from circumstellar radio CO lines. As in almost all 
miras (Lebzelter \& Hinkle \cite{LH02}) the velocity curve crosses the 
center-of-mass velocity near phase 0.4. 

S\,Car is noteworthy as it 
belongs to the group of intermediate population miras.  Other members of 
this group have been discussed in Lebzelter et al.\,(\cite{LHH99}).  These 
stars show shorter periods and earlier spectral types than classical miras. 
Their space motion reveals that this group belongs to an older population 
(e.g.\,Hron \cite{hron91}).  It was already noted in Lebzelter et al.\,
(\cite{LHH99}) that the velocity curves of these stars cannot be 
distinguished from those of classical miras, and it is therefore unlikely 
that these stars pulsate in a different mode. With the monitoring of 
S\,Car, we confirmed this result for another object.

Bumps in 
the rising branch of the light curve, appear occasionally in many miras. 
R\,Car falls in this category.
In addition, two stars commonly showing humps in their light curves have 
been observed previously, namely T\,Cas and T\,Cep (Hinkle et 
al.\,\cite{HSH84}). 
The question 
of the origin of these humps has been discussed by Lockwood \& Wing (\cite{lw}). They 
doubt the interpretation as a signature of an additionally excited first 
overtone mode in these stars (see e.g.\,the discussion in Barthes \& 
Tuchman \cite{BT94}), because of the stochastic nature of this phenomenon. 
Instead, they suggest that these humps result from an interplay between 
rising temperature and decreasing radius during this part of the light 
change.  In particular, they note that the temperature seems to continue 
its rise during the brief decline after the secondary maximum. This 
indicates that the radius becomes steadily smaller between the primary 
minimum and the primary maximum.  Such humps are also visible in the 
light curves produced by pulsation models for miras (Feuchtinger et al.\,
\cite{FDH93}; Ya'Ari \& Tuchman \cite{yt96}; Hofmann et al.\cite{hsw98}).


\subsection{A SRV with a long secondary period}

The velocity variations of W\,Nor, a SRV with a long secondary period, are 
clearly dominated by the long period. This has been found in several other 
stars as well (Hinkle et al.\,\cite{HLJF02}, Wood et al.\,\cite{WOK04}). 
Unfortunately, our time series is much too short to cover the whole 
pulsation cycle. Only a lower limit for the velocity amplitude can be 
given. However, this lower limit is already similar in value to the total 
velocity amplitudes observed in most other members of this group. V13 in 
47\,Tuc shows the largest measured amplitude of these stars in the near 
IR CO lines (12\,kms$^{-1}$).

\section{Conclusions}

With the help of near-infrared velocity curves we have investigated a variety
of phenomena at the border between miras and SRVs.  
Our sample of field stars with reliable Hipparcos parallaxes roughly follows 
the logP-$M_{K}$-sequences found in the LMC. Nearly all stars on sequence B (first 
overtone mode) show similar velocity amplitudes not exceeding a few 
kms$^{-1}$. Most of the objects pulsating in fundamental mode (sequence C) 
show much higher velocity amplitudes. Our result is in agreement with 
previous findings for LPVs in the globular cluster 47\,Tuc.

SRVs with periods around
300 days, and the large amplitude SRVs which occur over a wide range of
periods, all seem to be fundamental mode pulsators with smaller velocity
amplitudes than those of miras with similar periods.  They do not have the
characteristic discontinuity in the velocity curve found for miras, although at
least some of them show emission lines in their spectra.  It is
suggested that the smaller amplitude stars on sequence C have slightly higher
mass than miras adjacent to them on this sequence.

R\,Nor, a mira with a secondary maximum in its light
curve, has a velocity curve which does not show any evidence for a secondary 
maximum.  This shows that the
true pulsation period of the miras with secondary maxima is the interval
between secondary maxima.  The LMC analogs of stars like R\,Nor are
relatively massive (3--5 M$_{\odot}$) and luminous.  It is suggested that the
secondary maxima may be due to a resonance between the first overtone and
fundamental modes.

We have argued that as a star evolves up the AGB, it generally switches mode
from first overtone to fundamental mode. On the other hand, R\,Cen seems to be
rather on the way back: the history of the star's period decrease and light
curve change favour a switch from fundamental to first overtone mode. This may
be due to an ongoing Thermal Pulse, since the steady period change can be
interpreted as due to a decline in luminosity i.e. evolution {\it down} the AGB.

\begin{acknowledgements}
TL has been supported by the Austrian Academy of Science (APART programme),
PRW has been partially supported by Australian Research Council grant DP0343832
and FCF's research at Tennessee State University was partially funded by NASA grant 
NCC5-511 and NSF grant HRD-9706268. 
NOAO is operated by the Association of Universities for Research in Astronomy 
under cooperative agreement with the National Science Foundation.
Partly based on observations obtained at the Gemini Observatory, which is operated
by the Association of Universities for Research in Astronomy, Inc., under cooperative
agreement with the NSF on behalf of the Gemini partnership: the National Science Foundation
(US), the Particle Physics and Astronomy Research Council (UK), the National
Research Council (Canada), CONICYT (Chile), the Australian Research Council, CNPq (Brazil)
and CONICET (Argentina).
We acknowledge with thanks the variable 
star observations from the AAVSO International Database contributed by 
observers worldwide and used in this research. Especially we wish to acknowledge
the support for this and previous papers by AAVSO director Janet Mattei, who 
unfortunately deceased on March 22, 2004.
\end{acknowledgements}

\end{document}